\documentclass[aps, prd, amsmath, floats, floatfix, twocolumn,
superscriptaddress, nofootinbib, showpacs]{revtex4}
\usepackage{graphicx}
\usepackage{epsfig}
\usepackage{color}
\usepackage{soul}
\usepackage{url}
\usepackage{bm}         % bold math symbols
\usepackage{times}

\newcommand{\beq}{\begin{equation}}
\newcommand{\eeq}{\end{equation}}
\newcommand{\beqn}{\begin{eqnarray}}
\newcommand{\eeqn}{\end{eqnarray}}

\newcommand{\lo}{\mathrel{\raise.3ex\hbox{$<$}\mkern-14mu
    \lower0.6ex\hbox{$\sim$}}}
\newcommand{\go}{\mathrel{\raise.3ex\hbox{$>$}\mkern-14mu
    \lower0.6ex\hbox{$\sim$}}}

\usepackage{color}

\newcommand{\Caltech}{\affiliation{Theoretical Astrophysics 350-17,
    California Institute of Technology, Pasadena, California 91125, USA}}
\newcommand{\Cornell}{\affiliation{Center for Radiophysics and Space
    Research, Cornell University, Ithaca, New York, 14853, USA}}
\newcommand{\WSU}{\affiliation{Department of Physics \& Astronomy,
	Washington State University, Pullman, Washington 99164, USA}}
\newcommand{\CITA}{\affiliation{Canadian Institute for Theoretical 
    Astrophysics, University of Toronto, Toronto, Ontario M5S 3H8, Canada}}
\newcommand{\UofT}{\affiliation{Department of Astronomy \& Astrophysics,
    University of Toronto, Toronto, Ontario, M5S 3H5, Canada}}
\newcommand{\MUN}{\affiliation{Memorial University of Newfoundland, St. John's, NL, A1C 5S7, Canada}}

\usepackage{graphicx}% Include figure files
\usepackage{dcolumn}% Align table columns on decimal point
\usepackage{bm}% bold math
\usepackage{epsf}

\begin{document}

\title{First direct comparison of non-disrupting neutron star-black hole and binary black hole merger simulations}

\author{Francois Foucart} \CITA%
\author{Luisa Buchman} \Caltech%
\author{Matthew D. Duez} \WSU %
\author{Michael Grudich} \CITA \MUN 
\author{Lawrence E. Kidder} \Cornell %
\author{Ilana MacDonald} \CITA \UofT
\author{Abdul Mroue} \CITA
\author{Harald P. Pfeiffer} \CITA
\author{Mark A. Scheel} \Caltech %
\author{Bela Szilagyi} \Caltech %

\begin{abstract}
We present the first direct comparison of numerical simulations of neutron star-black hole and black hole-black hole
mergers in full general relativity. We focus on a configuration with non spinning objects and 
within the most likely range of mass ratio for neutron star-black hole systems ($q=6$). In this region of the parameter space,
the neutron star is not tidally disrupted prior to merger, and we show that the two types of mergers appear
remarkably similar. The effect of the presence of a neutron star on the gravitational wave signal is not only 
undetectable by the next generation of gravitational wave detectors, but also too small to be measured in the numerical
simulations: even the plunge, merger and ringdown signals appear in perfect agreement for both types of binaries.
The characteristics of the post-merger remnants are equally similar, with the masses of the final black holes agreeing
within $\delta M_{\rm BH}< 5 \times 10^{-4}M_{\rm BH}$ and their dimensionless spins within 
$\delta \chi_{\rm BH}< 10^{-3}$.
The rate of periastron advance in the mixed binary agrees with previously published
binary black hole results, and we use the inspiral waveforms to place constraints on the accuracy of our numerical
simulations independent of algorithmic choices made for each type of binary. 
Overall, our results indicate that non-disrupting neutron star-black hole mergers are exceptionally
well modeled by black hole-black hole mergers, and that given the absence of mass ejection, accretion disk formation,
or differences in the gravitational wave signals, only electromagnetic precursors could prove the presence
of a neutron star in low-spin systems of total mass $\sim 10M_\odot$, 
at least until the advent of gravitational wave detectors with a sensitivity comparable to that of the
proposed Einstein Telescope. 
\end{abstract}


\maketitle

\section{Introduction}
\label{sec:Intro}

Mergers of black holes and neutron stars are expected to be among 
the main sources of gravitational wave signals detectable by
the next generation of gravitational wave detectors (Advanced LIGO~\cite{Harry2010}, 
Advanced VIRGO~\cite{Acernese:2008}, KAGRA~\cite{Somiya:2012}), as well as by proposed 
'third generation' ground based detectors such as the Einstein Telescope~\cite{EinsteinTelescope}. 
Being able to associate a detected gravitational wave signal with a given type of 
binary system (binary black holes [BBH], binary neutron stars [BNS] or neutron star-black hole [NSBH] binary)
is an important way to gain useful insights into the formation mechanisms 
of compact binaries. However, even for the binary parameters for which the types of objects involved
have the largest effects, this is
generally a difficult task. In this paper, we will show through the first direct comparison 
of numerical simulations
of NSBH and BBH mergers that for configurations in which the tidal field of the black hole
is not strong enough to disrupt the neutron star, these two types of mergers are remarkably
similar - so much so that, even numerically, the differences in the gravitational
waveform, orbital evolution, and characteristics of the final remnant cannot be resolved.

The simplest method to associate a gravitational wave event with a given type of binary
relies on the determination of the mass of the compact objects,
as well as the assumption of a given maximum mass for neutron stars above which all 
observed objects are expected to be black holes. This works for objects which are 
either clearly too light to be black holes or too heavy to be neutron stars.
In the context of the Advanced LIGO detector, Hannam et al.~\cite{HannamEtAl:2013}
have recently studied which ranges of the measured chirp mass (the combination of the
masses of the two objects which is most accurately measured from the gravitational
wave signal) can be unambiguously associated with each type of binary. 
One limitation of this method, however, is that even if the
gravitational wave signals were strong enough for the individual masses to be
determined with high accuracy, compact objects would still be determined to
be either black holes or neutron stars based solely on our preconception
for the mass range in which neutron stars and black holes exist. It will never
allow us to determine, for example, whether a $2.5M_\odot$ object is a  
very massive neutron star or a very light black hole.

More direct evidence of the presence of a neutron star is the detection
of an electromagnetic counterpart to the gravitational wave signal,
either before the disruption of the neutron star (see e.g.~\cite{TsangEtAl:2012,PalenzuelaEtAl:2013,PaschalidisEtAl:2013}), 
after the merger (e.g. gamma-ray burst, kilonova, see~\cite{metzger:11,NissankeEtAl:2012}
for a review of these electromagnetic signals and of their detectability),
or, for BNS, during the ejection of relativistic material from the contact
layer between the two stars~\cite{KyutokuEtAl:2012}. The exact conditions under 
which premerger signals can be emitted are not, at this point, very well understood. 
The other signals
could be fairly common in binary neutron star mergers, although potentially
difficult to detect for most events (gamma-ray bursts
are expected to be strongly beamed, while isotropic counterparts are likely to be 
too faint to be detected for the farthest binaries observable by Advanced LIGO). 
For NSBH binaries, recent numerical simulations have shown that for
the most likely black hole masses ($M_{\rm BH}\sim 7M_\odot-10 M_\odot$~\cite{Ozel:2010,Kreidberg:2012}), 
the neutron star
will only be disrupted by the tidal field of the black hole for rapidly rotating black
holes~\cite{FoucartEtAl:2011,Foucart2012}. For lower mass black holes, tidal disruption
is significantly easier (see~\cite{Duez:2010,ShibataTaniguchi:2011} for reviews of numerical
simulations of NSBH mergers). Given that post-merger electromagnetic
signals cannot be emitted unless the neutron star is disrupted during the merger, this poses
strong constraints on the binary parameters for which we can hope to prove the presence of
a neutron star from post-merger electromagnetic observations.

Finally, measuring the effect of tides and, for
NSBH binaries, of the disruption of the neutron star on the gravitational
wave signal can allow us to distinguish black holes from neutron stars in merging binaries. 
For BNS, recent studies~\cite{damour:12,ReadEtAl2013} 
suggest that these effects will be measurable for a significant fraction of the Advanced LIGO 
events. For NSBH binaries, this would
only be possible for very low mass black holes~\cite{Lackey:2013}, or very close events
with rapidly rotating black holes (at most $\sim 1\%$ of the Advanced LIGO events
with black hole spin $\chi_{\rm BH}\sim 0.9$ and mass $M_{\rm BH}\sim 10 M_\odot$)~\cite{Foucart:2013a}.

For black holes in the most likely range of masses, tidal effects during inspiral
are expected to be orders of magnitude smaller~\cite{Hinderer2010}. The effects of the presence of a neutron
star on the plunge, merger, and characteristics of the post-merger remnant have not, however, been studied
in much detail. In terms of waveform analysis, the most complete study was performed 
by Lackey et al.~\cite{Lackey:2013}, where a large number of neutron star-black hole waveforms 
(all for black hole masses $M_{\rm BH}\leq 7M\odot$)
was presented and compared with analytical approximations to black hole-black hole waveforms. Their results provide us
with remarkable predictions for the effects of the equation of state of the neutron star on the gravitational wave signal emitted by low mass
neutron star-black hole mergers, and their detectability by Advanced LIGO and the Einstein Telescope. As for the expected characteristics of the final black
hole, a fit to the results of numerical simulations was recently proposed by Pannarale~\cite{Pannarale:2012}.
In both cases, the results resolve the effects of the presence of a neutron star in the regimes in which
these effects are the largest (low mass or rapidly spinning black holes). The various analytical approximations used to cover the parameter space
efficiently (numerical fits, use of approximate analytical waveforms) however dominate
the error in the regime of low-spin black holes in the most realistic mass range
(and, in the case of~\cite{Lackey:2013}, the Fisher matrix analysis applied to the waveforms breaks
down when estimating errors in the determination of the neutron star radius for high mass ratio systems,
as discussed in Sec.~\ref{sec:GW}). 

In this paper, we provide the first direct comparison between numerical simulations of NSBH and BBH mergers, 
and investigate in more detail the effect of the presence of a
neutron star during a merger with a nonspinning black hole of mass $M_{\rm BH}=6M_{\rm NS}\sim 8M_\odot$.
We show that, in every observable quantity that we considered, the two simulations are in remarkable agreement,
to very high accuracy. In particular, the periastron advance, merger waveform and post-merger characteristics
of the system are shown to be impossible to distinguish within the errors of the numerical simulations (and, a fortiori,
identical as far as gravitational wave detectors are concerned). Even for third generation gravitational wave 
detectors, the presence of a neutron star would influence the gravitational waveform mainly through 
a tiny phase shift accumulated during the inspiral, while the merger and ringdown would provide almost 
no additional information.

We also use these simulations to obtain a verification of the accuracy of our code during inspiral,
independent of the different algorithmic choices (gauge, grid setup,...) made in NSBH and BBH simulations. 
The inspiral waveforms at our highest numerical resolution agree within a phase accuracy 
$\delta \phi \lo 0.1\,{\rm rad}$, lower than the expected numerical errors (and significantly
larger than the dephasing due to tides within the neutron star). Our results also show that
the rate of periastron advance in NSBH simulations is compatible with
previously published results for BBH systems~\cite{LeTiec-Mroue:2011}, and we verify that tidal
effects in the neutron star are locally resolved and evolve as expected with the binary separation. 

\section{Numerical setup}
\label{sec:Methods}

We simulate BBH and NSBH binaries using the SpEC code, developed by
the SXS collaboration~\cite{SXSWebsite}. We use a
first-order~\cite{Lindblom:2007} generalized harmonic formulation of
Einstein's
equations~\cite{Friedrich1985,Garfinkle2002,Pretorius2005c,Lindblom:2007}
that are evolved with a multi-domain pseudospectral method. The
evolution region is divided into ``subdomains'' whose geometry (and
spectral basis) is adapted to the expected symmetries of the system in
each region (spheres close to the central objects and at large
distance from the binary, distorted cubes and cylinders
elsewhere). For NSBH mergers, the general relativistic equations of
hydrodynamics are evolved in conservative form on a separate finite 
difference grid, using high resolution shock-capturing
methods~\cite{Duez:2008rb}.  A more detailed description of the
numerical methods used for recent evolutions of NSBH binaries with the
SpEC code and of the grid structure used for the evolution of
Einstein's equations can be found in Foucart et
al.~\cite{Foucart:2013a}.

All simulations presented in this paper consider binaries of mass ratio $q=6$. Both
objects are initially nonspinning. The BBH results were first published as part of a
series of nonspinning BBH simulations by Buchman et al.~\cite{Buchman:2012dw}, where more details
can be found on the numerical setup used. The case considered
here covers 43 gravitational wave cycles (measured up to the peak of the dominant mode of the
gravitational waveform, $h_{22}$) at very low eccentricity (initially, $e=4\times 10^{-5}$).
The NSBH results are presented here for the first time. Although they represent the longest 
simulation of a NSBH merger published so far, they are still significantly
shorter than their BBH counterparts (25 cycles), and have a higher eccentricity 
($e<0.005$, see Sec.~\ref{sec:Acc} for a discussion of the eccentricity).
The neutron star fluid is modeled by a $\Gamma$-law equation of state
\beqn
P &=& \kappa \rho_0^\Gamma + \rho_0 T \\
\epsilon &=& \frac{1}{\Gamma-1} \frac{P}{\rho_0}
\eeqn
where $P$ is the pressure, $\epsilon$ the internal energy, $\rho_0$ the baryon density
and $T$ a quantity related to the temperature of the gas (i.e. $\rho_0 T$ is the thermal pressure).
We choose $\Gamma=2$ and $\kappa=92.12$, which for a star of gravitational mass $M_{\rm NS}=1.4M_\odot$
leads to a compactness $C_{\rm NS}=M_{\rm NS}/R_{\rm NS}=0.156$ (i.e. $R_{\rm NS}=13.3\,{\rm km}$)
\footnote{Throughout this article, we use the convention $G=c=1$, unless units are explicitly mentioned.}.
This radius is at the upper end of the current estimates for neutron star radii~\cite{Steiner2010,Guillot:2013wu}.
We generate constraint satisfying initial data using a spectral elliptic solver~\cite{Pfeiffer2003}, solving for
a quasi-equilibrium state through the iterative procedure described in Foucart et al.~\cite{FoucartEtAl:2008}.
Low-eccentricity orbits are obtained by evolving the binary for $\sim 3$ orbits, and using the measured
eccentricity of the simulation to obtain an improved guess for the initial orbital and radial
velocity of the binary, as described in Pfeiffer et al.~\cite{Pfeiffer-Brown-etal:2007}. One of these intermediate simulations
(with $e\sim 0.024$) was evolved for $\sim 8$ orbits to allow us to measure the advance
of the periastron (as this measurement cannot be made accurately on a low-eccentricity binary).

We run the NSBH inspirals at 3 different fixed resolutions (spectral resolution 
$N^{1/3}_{sp}=57,64,72$, and finite difference resolution 
$N^{1/3}_{fd}=100,120,140$).
Additionally, we run a 4th simulation using an adaptive choice of the spectral
resolution, where the number of basis functions over which we expand the solution in each of the
``subdomains'' forming our numerical grid is chosen adaptively, so that the truncation error of 
the spectral expansions of the metric components
and of their spatial derivatives are below $2\times 10^{-6}$ close to the black hole and neutron star,
and $2\times 10^{-4}$ far away from the compact objects. This offers significantly higher accuracy
throughout the inspiral, at a lower cost ($N^{1/3}_{sp}\sim 63$ during most of the
evolution, with a peak at $N^{1/3}_{sp}=73$ during the relaxation of the initial data). 
This additional simulation was performed to test that some unphysical effects visible at early times 
in the other simulations (see Sec.~\ref{sec:Acc}) converged away when the spectral resolution
was high enough (this simulation also used $N^{1/3}_{fd}=100$ on the finite difference grid, as
our tests show that, for this configuration, the finite difference resolution does not significantly 
influence the inspiral results).
The adaptive choice of the spectral resolution is also used for all 4 simulations during
the plunge/merger phases, 
with truncation error of $10^{-4},7\times 10^{-5},5\times 10^{-5}$ for the runs using fixed
resolution during the inspiral, and $10^{-4}$ for the simulation which used the adaptive method from
the beginning of the simulation
\footnote{
The merger in this case was only performed to extract the gravitational waveform at large
radii. Given the spectral and finite difference resolutions used during merger, the
post-merger results obtained from this simulation are not expected to be more accurate than
those of the low resolution simulation.
}.

It is also worth noting that different gauges have been used for each type of simulation.
In the generalized harmonic formulation of Einstein's equations, the coordinates $x^b$ evolve
according to
\beq
g_{ab} \nabla^c\nabla_c x^b = H_a(x,g_{ab})
\eeq
where $g_{ab}$ is the spacetime metric, and the $H_a$ are freely specifiable functions
of both the coordinates and the metric.  In the three ``fixed resolution'' runs, 
we fixed $H_a$ to its initial value in the coordinate frame 
comoving with the binary. For the last simulation (`adaptive' run) and the BBH simulations, 
we used the harmonic gauge $H_a=0$, with a smooth transition of $H_a$ from its initial value over a short
damping timescale $t_{\rm damp}=50M$ at the beginning of the simulation (where $M$ is the total mass
of the system). Numerical tests on the early part of the evolution of NSBH binaries
have shown that the harmonic gauge performs slightly better than
the frozen gauge, although that change makes a significantly smaller difference than
the use of the adaptive grid resolution. The harmonic gauge is also theoretically more satisfactory, as it
makes the evolution of the coordinates at late times independent of the initial configuration. 
Its only drawback is that, at times $t\sim t_{\rm damp}$, the coordinate radius of the apparent
horizon of the black hole decreases rapidly. The SpEC code requires the excision of a region
inside of the black hole, and maintains the boundary of that region within the apparent horizon
of the black hole. This is naturally more difficult if the apparent horizon is moving rapidly
on the grid, and made the use of the harmonic gauge impractical in NSBH simulations 
until recent improvements to the 
control system keeping the apparent horizon in place~\cite{Hemberger:2012jz}.
During mergers, $H_a$ is chosen according
to the ``damped harmonic'' prescription described by Szilagyi et al~\cite{Szilagyi:2009qz}.

Due to these differences in the choice of the gauge functions $H_a$, the coordinate evolution
of the binary can appear quite different for the various simulations. However, we will show
that all simulations agree remarkably well on more gauge-independent quantities.

\section{Accuracy}
\label{sec:Acc}

For the purpose of this comparison between BBH and NSBH results, the BBH simulations
can generally be considered as an accurate representation of the exact solution for BBH mergers.
Whether in terms of the phase and amplitude errors in the gravitational waveform, the
orbital eccentricity, or the properties of the final black hole, BBH simulations are indeed
at least an order of magnitude more accurate than their NSBH counterparts. A detailed
discussion of the numerical errors in the BBH results can be found in Buchman et al.~\cite{Buchman:2012dw}.
The error in the BBH and NSBH simulations are only of comparable magnitude when the uncertainty
due to the extrapolation of the waveform to spatial infinity and the representation of the waveform
as a finite sum of spherical harmonic modes induce errors of the same
order as the numerical error due to the use of finite resolution. Among the quantities discussed in
this section, this is only the case for the recoil velocity of the final black hole and the total
energy emitted in gravitational waves.

\begin{table}
\caption{
Post-merger properties of the black hole. $M_{\rm BH}^f$ is the final
mass of the black hole, $M=M_{\rm BH}+M_{\rm NS}$ is the
total mass of the system at infinite separation, 
$\chi_{\rm BH}=J_{\rm BH}/M_{\rm BH}^2$ is the dimensionless spin and
$v_{\rm kick}$ the kick velocity, as computed from the gravitational
wave emission. We also give the initial ADM energy of the system 
$E_{\rm ADM}^i$,  and the energy emitted in gravitational waves 
$E_{\rm GW}$ over the course
of the simulation. NsBh:L0, NsBh:L1 and NsBh:L2 are the 3 simulations
using fixed resolution (low, medium and high) during the inspiral, 
while NsBh:AMR used the adaptive choice of the spectral resolution.
}
\label{tab:simresults}
\begin{tabular}{|c||c|c|c|c|c|}
\hline
& $M_{\rm BH}^f/M$ & $\chi_{\rm BH}^f$ & $v_{\rm kick}(\,{\rm km/s})$ & $E_{\rm ADM}^i/M$ & $E_{\rm GW}/M$ \\
\hline
Bbh & 0.9855 & 0.3725 & 118 ($\pm 6$) & 0.9960 & 0.0104 \\ 
\hline
NsBh:L0 & 0.9832 & 0.3737 & 107 & 0.9953 & 0.0098 \\ 
NsBh:L1 & 0.9856 & 0.3727 & 105 & 0.9953 & 0.0098 \\ 
NsBh:L2 & 0.9854 & 0.3731 & 109 & 0.9953 & 0.0098 \\ 
NsBh:AMR & 0.9854 & 0.3726 & 109 & 0.9953 & 0.0099 \\
\hline
\end{tabular}
\end{table}

Although the NSBH simulations have significantly larger errors than the BBH simulations,
they are nonetheless very accurate by the standard of general relativistic simulations 
of compact mergers with matter. The mass and spin of the final black hole, listed in
Table~\ref{tab:simresults}, converge to an accuracy of $\sim 5\times 10^{-4}M$ for the mass and
$\sim 10^{-3}$ for the dimensionless spin (for the BBH case, these errors are $5\times 10^{-5}M$
and $10^{-4}$ respectively). The error in the determination of the recoil velocity imparted to the black
hole is mostly due to the extrapolation of the waveform to infinity, and is $\sim 10\,{\rm km/s}$, while the energy
emitted in gravitational waves is accurate to $10^{-4}M$ (for both the BBH and the NSBH simulations). 
Given that, at the end of the
simulation, there is no matter left outside of the black hole, we can also check
conservation of energy during the simulation, which requires
\beq
E^i_{\rm ADM} = E_{\rm GW} + M^f_{\rm BH}
\eeq
where $E^i_{\rm ADM}$ is the ADM mass of the binary at the initial time, $E_{\rm GW}$ is the
energy emitted in gravitational waves, and $M^f_{\rm BH}$ is the final mass of the black hole.
This equality is satisfied within the error in the final value of the black hole mass.

\begin{figure}
\includegraphics[width=0.95\columnwidth]{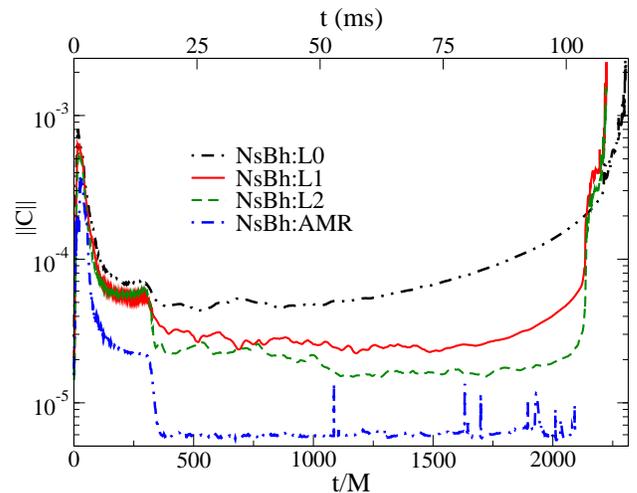}
\caption{Normalized constraint violation during the four NsBh inspirals.
$||C||$ is defined as in  Eq.(71) of~\cite{Lindblom2006}. The top axis (in $ms$)
assumes $M_{\rm NS}=1.4M_\odot$.}
\label{fig:GhCe}
\end{figure}

A commonly used measure of the accuracy of a numerical simulation in the generalized harmonic
gauge is the normalized constraint violation $||C||$, which measures the relative amplitude
of the modes violating the generalized harmonic gauge constraint, as well as additional constraints
introduced in the reduction of Einstein's equations to a set of first order differential equations.
An exact definition of $||C||$ can be found in Eq.(71) of~\cite{Lindblom2006}. 
In Fig.~\ref{fig:GhCe}, we show the evolution
of $||C||$ before the plunge of the neutron star into the black hole. From this figure, we can
clearly see the advantage of choosing adaptively the number of basis functions used in the spectral
decomposition of each of the ``subdomains'' forming our numerical grid: constraint violations are then
$\sim 5$ times smaller than for the high-resolution run, while the number of grid points is nearly
the same as for the medium-resolution run. Additionally, at early times ($t<300M$), the fixed
resolution runs are not in the convergent regime, and convergence
remains slow until $t\sim 1000M$.

\begin{figure}
\includegraphics[width=0.95\columnwidth]{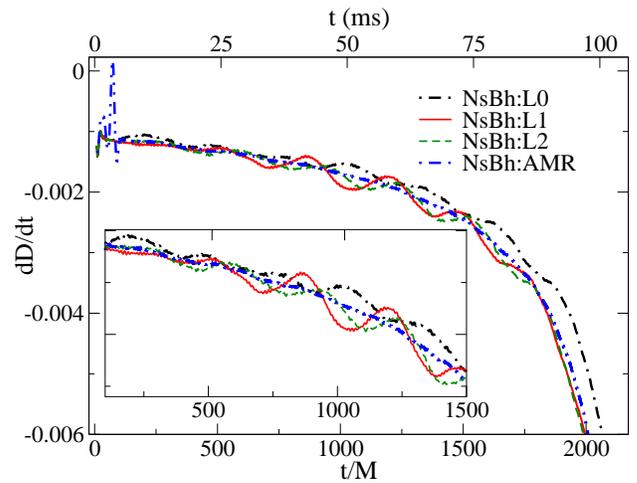}
\caption{Evolution of the time derivative of the coordinate separation between the center of the
black hole and the center of the neutron star for the four NsBh simulations. The inset zooms on the
early time behavior.}
\label{fig:ecc}
\end{figure}

Most global quantities are largely unaffected by these issues. In particular, errors in the
final characteristics of the system presented in Table~\ref{tab:simresults} are mostly
determined by the resolution during the merger, where numerical errors are significantly
larger (at the highest resolution, $||C||\sim 0.001-0.01$ during merger). However, a subtle
effect of the numerical error at early times in the ``fixed resolution'' simulations is the
evolution of the eccentricity. Fig.~\ref{fig:ecc} shows the evolution of the time derivative
of the coordinate distance $D$ between the centers of the compact objects. The three
``fixed resolution'' simulations appear to converge towards the simulation using adaptive
grid choices, but convergence is fairly slow, and the numerical error largely appears
as a growth of the eccentricity, from $e<0.001$ at early times to $e\sim0.005$ for $t\sim 1000M$.
In the more accurate simulation using adaptive grid choices, this effect is entirely
removed. At the level of accuracy obtained in these simulations, this effect is enough
to affect the phase of the gravitational waveform and, in particular, the time at which
the neutron star plunges into the black hole. This is visible in the fact that, despite
visible differences in their evolution during the inspiral, the two highest ``fixed
resolution'' simulations appear to agree extraordinarily well on the merger time --
an agreement which is clearly accidental once the entire evolution is considered.
Such accidental agreement is a strong warning as to the dangers of estimating
numerical errors by solely comparing numerical simulations at two different resolutions
--- a procedure which, in this case, would lead us to significantly underestimate
the error on the plunge time (and, consequently, on the gravitational wave phase 
at merger).

\begin{figure}
\includegraphics[width=0.95\columnwidth]{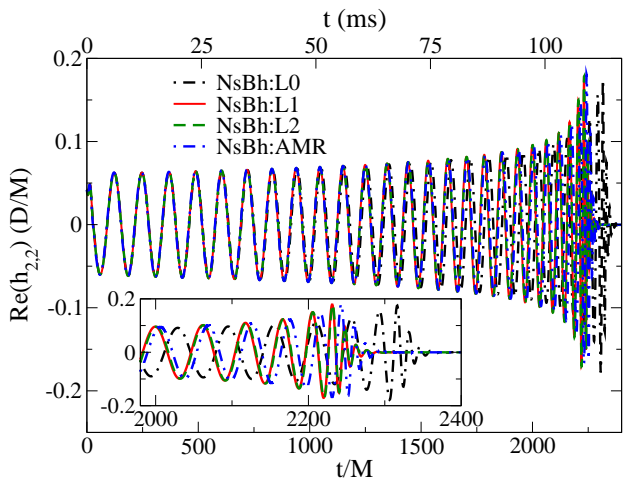}
\caption{Dominant (2,2) mode of the gravitational waveform for the NSBH system 
at low, medium and high resolutions, as well as for the `adaptive' run. 
The insert zooms on the time of merger.}
\label{fig:gwconvnomatch}
\end{figure}

\begin{figure}
\includegraphics[width=0.95\columnwidth]{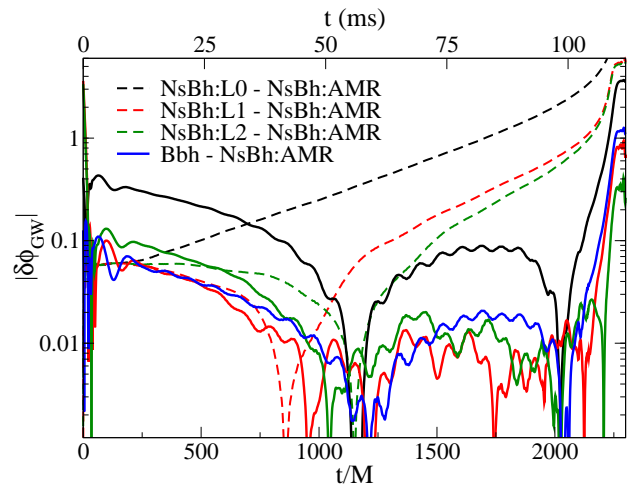}
\caption{Phase error in the (2,2) mode of the gravitational waveform. 
 The dashed curves show the difference between the ``fixed resolution''
simulations and the adaptive simulation without
time and phase shifts, while the solid curves
show the same phase differences, but after matching the waveforms in the
interval $675<t/M<2175$. The solid blue curve shows the difference between
the adaptive simulation and the BBH results (with arbitrary shifts).}
\label{fig:gwphaseerror}
\end{figure}

The error on the dominant (2,2) mode of the strain $h_{2,2}(t)$ can be estimated
from Fig~\ref{fig:gwconvnomatch}, which shows the real part of $h_{2,2}$, and
Fig.~\ref{fig:gwphaseerror}, which shows the difference in the phase of 
$h_{2,2}$ between the ``fixed resolution''
simulations and the ``adaptive'' simulation both with and without allowing the
results to be shifted by an arbitrary time and phase shift. In all cases, the 
waveforms are obtained using $2^{\rm nd}$ order polynomial extrapolation
from the waveforms measured at 20 finite radii in the interval $[100M-275M]$,
to spatial infinity, using techniques developed in~\cite{Boyle-Mroue:2008}.
For the two highest fixed resolutions, we see the effect of the coincident
agreement in the plunge time in an apparently small error in the gravitational
wave phase. A better estimate of the error is given by the difference with the
adaptive simulation, which is $\sim 0.1{\rm rad}$ during inspiral and $\sim 1{\rm rad}$
at merger (for comparison, the error in the BBH simulation is $\lesssim 0.01{\rm rad}$
during inspiral and $\lesssim 0.3{\rm rad}$ at merger). 
In fact, given that the expected tidal dephasing due to finite size effects
in this system is well below the numerical error (Post-Newtonian estimates
indicate that it should be $\lesssim 0.1{\rm rad}$  without any time/phase shift 
for $t<2000M$~\cite{Hinderer2010,Vines2011,damour:12}, 
and more than an order of magnitude smaller with the shifts), the phase
difference between the NSBH simulations and the BBH simulation during the inspiral
is an equally valid estimate of the numerical error. And indeed,
the difference between the phase of the gravitational waveform of the BBH simulation
and that of the ``adaptive'' NSBH simulation is of the same order, and of the opposite sign,
as the difference between the most accurate ``fixed resolution'' NSBH simulations and 
the adaptive one.
Considering that at high resolution our code is only expected to be $\sim 2^{\rm nd}$ order convergent, 
obtaining the order of magnitude increase in accuracy which would be required to resolve tidal
effects during the inspiral of such a NSBH binary is not a realistic objective at this point
\footnote{Obtaining higher order convergence in GR-Hydro codes is however possible, and has in fact been
recently tested in the study of BNS systems by Radice et al.~\cite{Radice:2013cba}},
nor particularly interesting considering that these effects cannot be resolved by the
next generation of gravitational wave detectors (see Sec.~\ref{sec:GW}). Studying these
effects in lower mass ratio binaries (or BNS systems) is a more realistic objective from
both a numerical and observational point of view.
 
Finally, the error in the amplitude $A$ of the waveform is $\delta A /A \sim 2\%$,
for both the time domain waveform $h_{2,2}(t)$ and its Fourier transform $\tilde h_{2,2}(f)$
up to high frequency ($f\lesssim 2\,{\rm kHz}$),
which will allow us to compare the BBH and NSBH spectra with high accuracy. Higher order
modes have slightly larger relative errors ($\sim 5\%$ for the first two subdominant modes,
$h_{3,3}$ and $h_{2,1}$).

\section{Results}
\label{sec:nr}

\subsection{Periastron Advance}
\label{sec:periastron}

The trajectories of compact objects during an inspiral are intrinsically gauge-dependent,
and thus difficult to compare among simulations that do not use the exact same gauge
prescription. There is however one important general relativistic effect acting on the
trajectories which, for nearly circular orbits and a large class of coordinate systems, is
independent of these gauge choices: the rate of periastron advance of the orbit, i.e.
the change in the angular location of the periastron between two periastron passages.
Explaining the periastron advance of the orbit of Mercury, first observed by Le Verrier
in 1859, was one of the early successes of the theory of general relativity~\cite{1915SPAW...47..831E}.
While this effect is small for Mercury ($\sim 43''$ per century), it becomes very significant
for compact objects close to mergers. In the context of BBH systems, periastron advance was
measured for a number of nonspinning BBH systems by Le Tiec et 
al.~\cite{LeTiec-Mroue:2011}, and shown to match remarkably well the predictions of the effective
one-body (EOB) formalism and of the self-force
theory (at least if the expansion in the mass ratio used in the self-force formalism is done using 
the symmetric mass ratio $\nu=m_1 m_2 / (m_1+m_2)^2$ instead of $q=m_1/m_2$). In this section, 
we follow the same procedure to provide a measurement of the periastron advance for
a NSBH binary of mass ratio $q=6$ and orbital eccentricity $e=0.024$, evolved for 8 orbits
from an initial orbital frequency $\Omega M=0.027$.

\begin{figure}
\includegraphics[width=0.95\columnwidth]{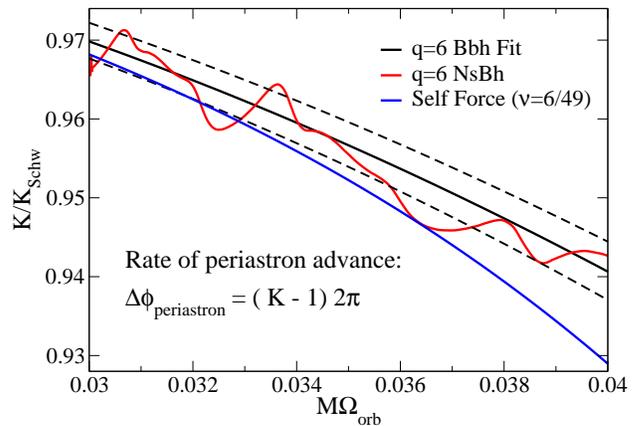}
\caption{Measurement of the rate of periastron advance as a function of the orbital frequency
for a NSBH system of mass ratio $q=6$,
compared with a fit valid for BBH systems~\cite{LeTiec-Mroue:2011}, and the results
from first-order self-force calculations~\cite{Barack:2010ny,LeTiec-Mroue:2011}. All values are normalized by the 
rate of periastron advance for a point particle around a Schwarzschild black hole.
The dashed lines show the uncertainty in the fit to the BBH results.}
\label{fig:periastron}
\end{figure}

We define the frequency of radial oscillations $\Omega_r=2\pi/P$ and the
average orbital frequency 
\beq
\Omega_\phi =\frac{1}{P}\int_0^P \frac{d\phi}{dt} = K \Omega_r,
\eeq
where $P$ is the period of radial oscillations of the orbit and $\phi$ the orbital phase
of the binary. After each radial period, the periastron will advance by an angle
$\Delta \Phi = (K-1) 2\pi$. To measure $K(t_0)$, we choose a time interval
covering 1.5 orbital periods of the binary centered on the time $t_0$\footnote{We verified that
the measured value of $K$ is insensitive to the choice of the fitting interval by using
2 and 2.5 orbital periods instead, and checking that the results are only modified at the
level of the small scale oscillations visible in Fig.~\ref{fig:periastron}.}, 
and fit the orbital
frequency $\Omega(t)$ (measured from the coordinate trajectories) to the model
\beq
\Omega(t) = p_0 (p_1-t)^{p_2} + p_3 \cos{\left(p_4 + p_5 (t-t_0) + p_6 (t-t_0)^2 \right)}.
\eeq
We then extract the average orbital period $\Omega_\phi=p_0 (p_1-t_0)^{p_2}$ and the period
of radial oscillations $\Omega_r=p_5$ (as the periodic variation of $\Omega(t)$ is due
to the eccentric motion of the binary). The constant $K$ is then 
$K_{\rm NSBH}=\Omega_\phi/\Omega_r$.

In Fig.~\ref{fig:periastron}, we show the result of this measurement, normalized by
the value for a point particle around a Schwarzschild black hole 
$K_{\rm Schw}=(1-6x)^{-1/2}$ (where $x=(M \Omega_\phi)^{2/3}$ is the Post-Newtonian
expansion parameter)~\cite{Damour-Schafer:1988,Cutler-et-al1994}. Figure~\ref{fig:periastron}
also shows the predictions of first order self-force calculations~\cite{Barack:2010ny,LeTiec-Mroue:2011}, 
as well as the measured value for BBH systems. 
For the latter,
we use the fit to the numerical data provided by Le Tiec et al.~\cite{LeTiec-Mroue:2011},
\beq
K_{\rm BBH}=K_{\rm Schw}\left(a_0 + a_1 (M \Omega_\phi) + a_2 (M \Omega_\phi)^2 \right).
\eeq
For a mass ratio $q=6$, the best-fit coefficients are $a_0=0.9890$, $a_1=1.071$, 
and $a_2=-57.0$ for a BBH system.

The measurement of $K_{\rm NSBH}$ provided here is clearly in good agreement with the BBH results,
with 
\beq
|K_{\rm NSBH}-K_{\rm BBH}|/K_{\rm BBH}\leq 0.5\%.
\eeq
This is comparable to the numerical error in the determination of $K_{\rm NSBH}$ (which can be estimated from
the oscillations of $K_{\rm NSBH}$), and becomes better than the accuracy of the self-force predictions at high 
frequencies ($\Omega_{\rm orb}M \go 0.036$).
We also checked separately that a similar agreement was observed for a shorter simulation at mass ratio
$q=5$ (6 orbits from an initial orbital frequency $\Omega M=0.028$, with eccentricity $e\sim 0.025$). 
The finite size of the neutron star has no measurable effect on the radial oscillations of the binary.

\subsection{Tidal distortion}

\begin{figure}
\includegraphics[width=0.95\columnwidth]{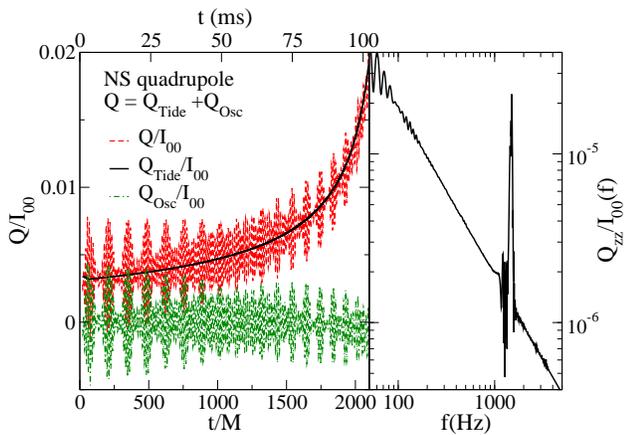}
\caption{Tidal quadrupole during the inspiral of the ``adaptive'' run. {\it Left:} Amplitude of the tidal
quadrupole $Q$, best fit $Q_{\rm Tide}$ to the expected evolution of the equilibrium tide (proportional to $d^{-3}$,
where $d$ is the binary separation), and residual $Q_{\rm osc}$ from that fit. {\it Right:} Fourier transform of 
$Q_{zz}/I_{00}$, composed of a power-law component (equilibrium tide) and an oscillatory component
at $1.5\,{\rm kHz}$ (assuming $M_{\rm NS}=1.4M_\odot$).}
\label{fig:Q}
\end{figure}

The evolution of the tidal distortion of the neutron star also follows closely the lowest order predictions
for the effect of the tidal field of the black hole. As in~\cite{Foucart:2013a}, we compute
the quadrupole moments of the neutron star in the coordinate frame corotating with the binary,
\beq
Q_{ij}=\int \rho \left(x_i x_j - \frac{1}{3} \delta_{ij}r^2\right) dV
\eeq
(where $x^i$ is a coordinate system whose origin is the center of mass of the neutron star, and oriented
so that its $x$-axis points away from the black hole and its $z$-axis is parallel to the orbital angular
momentum), as well as the second moment of the density
\beq
I_{00}=\int \rho r^2 dV
\eeq
where $\rho=\sqrt{g}W\rho_0$ and $W$ is the Lorentz factor of the fluid. We then extract the tidal
part $Q$ of the quadrupolar distortion from the coordinate distortion due to the Lorentz boost given
to the neutron star using the approximate method described in~\cite{Foucart:2013a}. The results
are shown in Fig.~\ref{fig:Q}. The time dependence of $Q/I_{00}$ clearly has two main components:
the expected growth of the equilibrium tide as the neutron star gets deeper into the tidal field of the black hole,
which is proportional to $d^{-3}$ (where $d$ is the coordinate separation of the binary), and a slowly
damped oscillatory component, due to the excitation of resonances in the neutron star in the imperfect
initial data. To separate the two, we fit $Q_{\rm Tide}/I_{00}=\alpha d^{-3}$ to find the equilibrium
tidal component, and subtract $Q_{\rm Tide}$ from the measured $Q$ to obtain the oscillatory component.
The equilibrium component clearly captures the expected long term evolution of $Q$. 
In a previous paper~\cite{Foucart:2013a}, we had also observed a surprisingly good agreement between 
numerical results and the lowest order theoretical prediction
\beq
\frac{Q}{I_{00}} \sim 2 k_2 R_{\rm NS}^5 \frac{M_{\rm BH}}{I_{00}d^3}
\eeq
(where $k_2$ is the tidal Love number of the neutron star, which for the equation
of state used here was computed by Hinderer~\cite{Hinderer:2008}, and $I_{00}$ is
computed for an isolated neutron star).
This is not the case for the simulations presented here, where the numerical
value of $Q/I_{00}$ is $\sim 50\%$ larger than the theoretical
value. This is not particularly surprising: the normalization of $Q/I_{00}$ is expected to be
gauge-dependent (e.g. it depends on the definition of the coordinate distance $d$), and the
order-of-magnitude agreement found here is theoretically more reasonable than the nearly exact
(and probably coincidental) agreement found in~\cite{Foucart:2013a}. 

The high-frequency oscillations shown here were not resolved in~\cite{Foucart:2013a}, where
$Q$ was only computed at a small number of discrete times. They simply appeared as a random
error on the measured value of $Q$. The improved sampling rate used in this work allow us
to clearly show that these deviations are not just due to errors in the method used to extract 
$Q$, but are instead due to the ringing of a stellar mode at $f\sim 1.5\,{\rm kHz}$ 
(see the right panel of Fig.~\ref{fig:Q}, in which we show the Fourier transform of $Q_{zz}/I_{00}$,
the component of $Q_{ij}$ which is not affected by the transformation between
the inertial and corotating coordinate systems). 
The amplitude of these oscillations decreases over time, albeit
very slowly (by about a factor of 2 over the course of the entire simulation). This mode
is excited in the imperfect quasi-equilibrium initial data, and rings without much 
dissipation during the inspiral. Despite the fact that the oscillatory part of $Q$ is
initially of the same amplitude as the equilibrium tide, it should have a negligible
effect on the evolution of the orbit and the emitted gravitational waves because,
as opposed to the tidal quadrupole, it does not efficiently couple to the orbital
quadrupole.

\subsection{Merger dynamics and final remnant}

Agreement in the rate of periastron advance between BBH and NSBH simulations, albeit reassuring,
is not overly surprising considering that effects of the finite size of the neutron
star during the inspiral (e.g. tides) are known to have only a small influence on the orbital evolution of
a binary of mass ratio $q=6$. It is only during the plunge and merger that we would expect 
larger differences to occur, both in the properties of the final black hole and in
the gravitational wave spectrum at high frequency. Instead, we find that their observable
properties (gravitational waves, characteristics of the post-merger remnant,...) 
are surprisingly similar.

\begin{figure}
\includegraphics[width=0.95\columnwidth,bb=100 340 900 820]{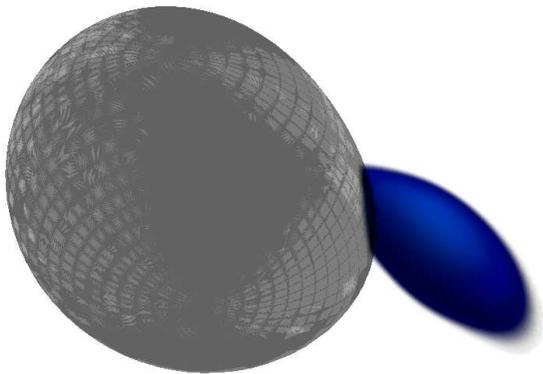}
\caption{Matter distribution (in blue) and location of the apparent horizon (in grey) after $5\%$ of the neutron star
material has been accreted onto the black hole. Even at merger, the neutron star remains remarkably compact.}
\label{fig:Merger}
\end{figure}

For a binary of mass ratio $q=6$ and a non spinning black hole, we expect the neutron
star to reach the innermost stable circular orbit (ISCO) of the black hole before
tidal effects cause the star to overflow its Roche lobe (thus causing unstable mass
transfer onto the black hole, and the disruption of the neutron star): for the neutron
star of compactness $C_{\rm NS}=0.156$ considered here, a dimensionless black hole 
spin $\chi_{\rm BH}\gtrsim 0.6$ is expected
to be required for disruption to occur before the plunge~\cite{Foucart2012}.
But even during the plunge, the neutron star is largely unaffected
by the tidal field of the black hole: Fig.~\ref{fig:Merger} shows the matter distribution as material 
from the neutron star
begins to cross the apparent horizon of the black hole. From this gauge-dependent visualization,
it appears that the distortion of the neutron star is minimal even at the point at which
the largest tidal effects are expected: deviations from spherical symmetry
are barely larger in the neutron star than in the black hole.

Another way to see this lack of distortion of the neutron star is to look at the accretion of
material onto the black hole: the entire neutron star crosses the apparent horizon of the black
hole within $0.5{\rm ms}$ - only a few times the light-crossing time across the undisturbed neutron
star.

But the most striking agreement between the BBH and the NSBH results is probably in the characteristics of
the resulting black holes, summarized in Table~\ref{tab:simresults}: the masses of the final black holes agree
within an accuracy of $\delta M_{\rm BH}^f \lo 5\times 10^{-4}M$, and their spins within 
$\delta \chi_{\rm BH}^f \lo 0.001$!
These constraints are more than an order of magnitude tighter than existing predictions based on numerical fits
to previous simulations of NSBH and BBH mergers~\cite{Pannarale:2012}.
Even the velocity kicks given to the remnants are in good agreement between the NSBH and BBH simulations, with
$v_{\rm kick}\sim 100-125\,{\rm km/s}$ for all cases\footnote{Note that the error in the
computation of $v_{\rm kick}$ is due in about equal amounts to the finite resolution used in the simulation
and to the procedure used to extrapolate the gravitational waveform to infinity and measure from it
the linear momentum emitted by the system.}. 

For non-spinning objects, in the most likely range
of black hole masses, the merger and post-merger evolution of NSBH binaries will thus look exactly identical to 
their BBH counterparts: no accretion disk will be formed,
no matter will be ejected, and differences in the properties of the final black hole will be negligible.

This agreement occurs despite the fact that the two types of mergers have different topologies. The event horizon
of a NSBH merger, which cannot be directly computed in our numerical simulations, should form a single worldtube
instead of the merging event horizons (``pants'' diagram) typical of BBH mergers. The marginally trapped surfaces
(apparent horizons) which are followed by our code also have different geometries. In NSBH mergers, we observe
a continuous evolution of the apparent horizon of the initial black hole towards the apparent horizon
of the final black hole. In BBH mergers, the two apparent horizons of the initial black holes coexist on some spacelike
hypersurfaces with an outer apparent horizon surrounding them (including on some hypersurfaces used as constant time
slices in our numerical evolutions). From a geometrical standpoint, NSBH and BBH mergers are thus very different events
--- but our results show that these differences do not significantly affect the observable properties of the merger.

\subsection{Gravitational Waveforms}
\label{sec:GW}

We have just seen that, for the parameters considered here, NSBH and BBH mergers are remarkably similar in terms
of orbital evolution, merger dynamics, and local properties of the post-merger remnant. The main observable counterpart
to the merger of a NSBH binary, however, is the gravitational wave signal emitted during its long inspiral and
eventual merger. It would thus be natural to expect that differences in their gravitational waveforms 
would be the easiest way to tell NSBH and BBH systems apart. In this section, we will see that although 
this is indeed likely to be the case, it remains an extremely challenging task which for black holes
of mass $M_{\rm BH}\go 7M_\odot$ is probably
beyond the reach of the upcoming advanced gravitational wave detectors.

Recent simulations have slowly made this more apparent: if for binary neutron stars the effect
of the finite size of the neutron star is expected to be detectable for a significant fraction
of Advanced LIGO events~\cite{damour:12,ReadEtAl2013} (assuming that those finite size effects can be accurately modeled,
which remains an important area of research), things become significantly more complicated once
NSBH mergers are considered. Finite size effects can be marginally resolved in low mass
systems~\cite{Lackey2011,Lackey:2013} or, in the most likely range of black hole masses, for at most a few percents 
of the events
with high black hole spin~\cite{Foucart:2013a,Lackey:2013}. This additional difficulty comes from the fact
that tidal effects during the inspiral are significantly reduced for asymmetric mass ratios, while the disruption of the
neutron star (which causes a sharp cutoff in the gravitational wave emission) occurs at frequencies slightly above 
the main Advanced LIGO band ($f_{\rm cut}\go 1.5\,{\rm kHz}$). Unless the astrophysical population of NSBH binaries
is particularly favorable for upcoming gravitational wave detectors, it is thus likely that finite size effects
in NSBH binaries will only be detectable with an improved detector such as the proposed Einstein Telescope.

The details of the gravitational wave signals in the case of NSBH mergers in which the neutron star does not disrupt,
and in particular their differences with signals from BBH mergers, remained until now largely unexplored. The best comparison
to date, by Lackey et al.~\cite{Lackey:2013}, resolves finite size effects in the low mass ratio limit, as well as the
disruption of the neutron star when it occurs.  However, the $\sim 30\%$ error expected in the modeled gravitational wave spectrum at disruption,
and the uncertainties in the phenomenological waveforms used as references for the BBH cases, do not allow for accurate direct
comparisons of BBH and NSBH results in the regime that we consider here ($q=6$, non spinning).
\footnote{See also~\cite{PannaraleEtAl2013} for a more accurate model of the amplitude of the gravitational wave signal emitted by non-spinning
NSBH binaries, but without any phase information.}
We do not, of course, expect larger differences for these
non-disrupting cases. However, it was a priori unclear how much more difficult the detection of finite size effects would
be for a non-spinning black hole around the peak of the black hole mass function. Indeed, even though the neutron
star was expected to reach the ISCO largely undisturbed, the ISCO frequency is still relatively low 
($f_{\rm ISCO}^{\rm GW}\sim 600\,{\rm Hz}$ for a non-spinning black hole of mass $M_{\rm BH}\sim 8.4M_\odot$ 
and for $q=6$, according to first order self-force calculations~\cite{Barack:2010tm}). 
The plunge, merger and ringdown thus represent a larger fraction of the detectable signal, starting at a frequency well below
the cutoff frequencies observed in disrupting binaries.

\begin{figure}
\includegraphics[width=0.95\columnwidth]{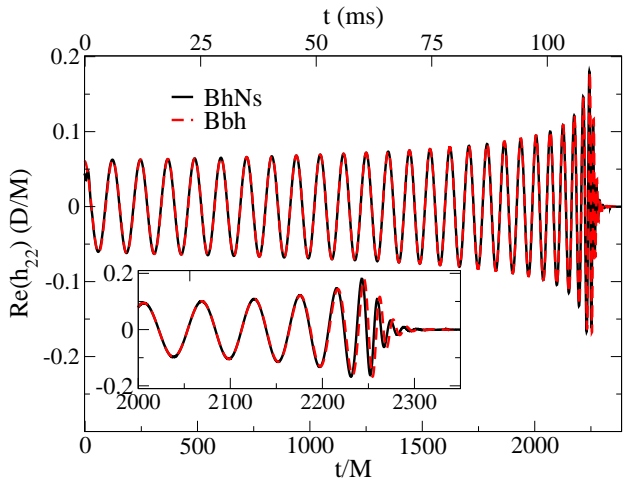}
\caption{Dominant (2,2) mode of the gravitational waveform for the NSBH (`adaptive' run)
and BBH mergers. The insert zooms on the time of merger. A time and phase shifts have been
applied to the BBH waveform in order to minimize the phase difference with the NSBH results 
in the interval $675<t/M<2175$.}
\label{fig:gwBbhVsBhNs}
\end{figure}

\begin{figure}
\includegraphics[width=0.95\columnwidth]{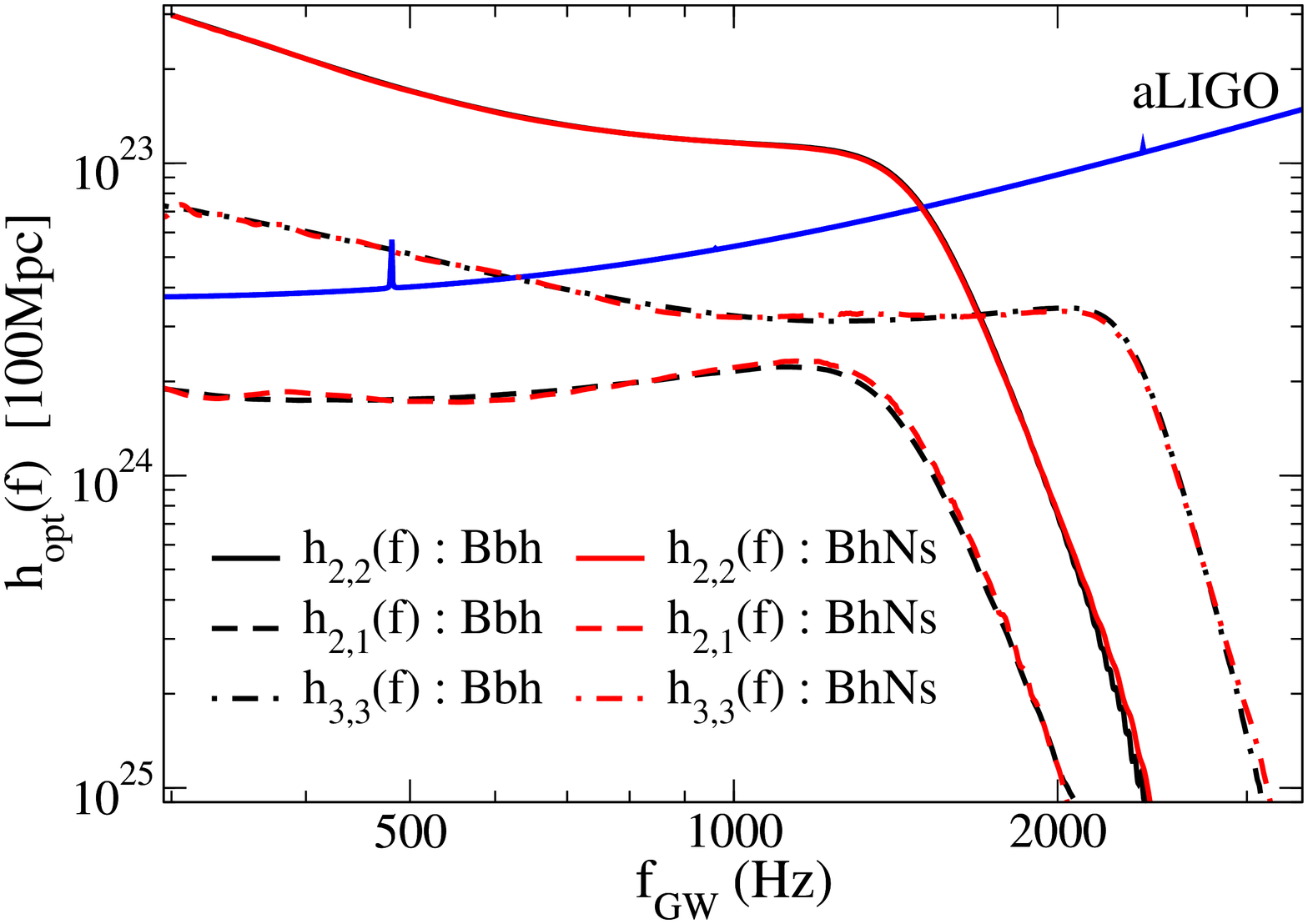}
\caption{Amplitude of the gravitational wave spectrum for optimally oriented binaries located at $100\,{\rm Mpc}$.
The aLIGO curve is the Zero-Detuned High Power noise curve of Ref.~\cite{Shoemaker2009}. BBH results are shown in
black and NSBH results in red for the dominant mode ($l=2$, $m=2$, solid curves), and the two largest subdominant modes
($l=2$, $m=1$, dashed curves and $l=3$, $m=3$, dash-dotted curves). The amplitude $h_{\rm opt}$ is defined as
$h_{{\rm opt},lm}(f)=\sqrt{5/4\pi}f^{1/2}h_{lm}(f)$.}
\label{fig:Spectra}
\end{figure}

Alas, even after the neutron star reaches the ISCO the signal remains remarkably devoid of any imprint of the presence of a
neutron star. Fig.~\ref{fig:gwBbhVsBhNs} shows the real part of the strain in the time domain for the BBH and NSBH mergers. 
The difference between the two waveforms is within the numerical error of the NSBH simulations.
Fig.~\ref{fig:Spectra} shows the spectrum of the dominant ($l=2$, $m=2$) mode of the gravitational wave signal 
as seen by an optimally oriented observer $100\,{\rm Mpc}$ away, for both the BBH and NSBH systems. The two cannot be distinguished
at the $\sim 2\%$ accuracy level of our simulations, even at frequencies $f\sim 1-2\,{\rm kHz}$ well beyond the ISCO frequency.
The same is true of the two largest subdominant modes, also shown in Fig.~\ref{fig:Spectra}.

From these waveforms, we can directly obtain an upper bound on the distance at which a gravitational wave detector would be
able to observe the difference between a NSBH and a BBH merger, if it was only looking at the time frame covered by the
numerical simulations. To do so, we define the difference $||\delta h||$ between two waveforms $h_1$ and $h_2$ as
\beq
||\delta h|| = \min_{\Delta t, \Delta \phi}\left(\sqrt{\langle h_1-h_2,h_1-h_2 \rangle}\right)
\eeq
where the product $\langle g,h \rangle$ is given by
\beq
\langle g,h \rangle =  2 \int_0^{\infty} df
\frac{\tilde{g}^*(f)\tilde{h}(f) + \tilde{g}(f)\tilde{h}^*(f)}{S_n(f)}
\label{eq:gdoth}
\eeq
and we have applied to $h_2(t)$ a time shift $\Delta t$ and phase shift $\Delta \phi$
minimizing $||\delta h||$.
Here $\tilde{g}(f)$ and $\tilde{h}(f)$ are the Fourier transforms of
$g(t)$ and $h(t)$, and $S_n(f)$ is the one-sided power
spectral density of the detector's strain noise, defined as 
\beq
S_n(f) = 2 \int_{-\infty}^{\infty} d \tau \, e^{2 \pi i f \tau}\,
C_n(\tau)\;,\qquad f>0,
\eeq
where $C_n(\tau)$ is the noise correlation matrix for zero-mean,
stationary noise. Taking for $S_n(f)$ the Zero Detuned High Power spectrum from~\cite{Shoemaker2009},
which is the design sensitivity of Advanced LIGO, and limiting the integral in Eq.~(\ref{eq:gdoth})
to frequencies $f>0.3\,{\rm kHz}$, we find the differences $||\delta h||$ listed in Table~\ref{tab:mismatch}.
If we neglect degeneracies between the effect of the finite size of the neutron star and other parameters of the binaries,
an approximate condition for the difference between two waveforms to be detectable is
$||\delta h||>1$~\cite{Lindblom2008}. Accordingly, our results show that, at best, differences in the merger
waveforms would be observable by Advanced LIGO for optimally oriented binaries located within $\sim 10\,{\rm Mpc}$ of the 
detector -- or about once in a million events. Clearly, the high frequency portion of the waveform alone will be
of no use to an observer attempting to determine whether a gravitational wave signal comes
from a BBH or a NSBH binary.

\begin{table}
\caption{
Differences $||\delta h||$ between numerical waveforms computed for frequencies 
above $0.3\,{\rm kHz}$. All values are computed for the Zero-Detuned High Power
noise curve of the AdvLIGO detector~\cite{Shoemaker2009}, and an optimally
oriented source located at $100\,{\rm Mpc}$.
}
\label{tab:mismatch}
\begin{tabular}{|c||c|c|c|c|}
\hline
vs & Bbh & NsBh:L0 & NsBh:L1 & NsBh:L2 \\
\hline
Bbh & -- & 0.17 & 0.08 & 0.06 \\
NsBh:L0 & 0.17 & -- & 0.22 & 0.22 \\
NsBh:L1 & 0.08 & 0.22 & -- & 0.04 \\
NsBh:L2 & 0.06 & 0.22 & 0.04 & --\\
\hline
\end{tabular}
\end{table}

Of course, in practice, gravitational wave detectors will not rely solely on the merger waveform. Most of the
signal-to-noise is accumulated at lower frequencies during the inspiral. Finite size effects during the inspiral
have been estimated in the Post-Newtonian formalism~\cite{Flanagan2008,Hinderer2010,Vines2011,damour:12}. 
They cause NSBH and BBH systems to accumulate a small phase
shift over time. For the binary parameters considered here, this phase shift is much 
below the phase accuracy
of our code (see Sec.~\ref{sec:Acc}), and the measured phase difference between our numerical BBH and NSBH waveforms
before merger (see Fig.~\ref{fig:gwphaseerror}) is mostly a measure of the accuracy of our code.
Nevertheless, tidal dephasing during the binary inspiral is the easiest way to measure finite size effects
in the gravitational waveform for a $q=6$, non spinning system. As in~\cite{Foucart:2013a}, we can estimate
the difference $||\delta h||$ between the BBH and NSBH waveforms by attaching to the numerical merger a Post-Newtonian
inspiral, with or without tidal terms. For the Post-Newtonian waveform, we use the Taylor T1 expansion, 
which matches numerical results very well for BBH at $q=6$~\cite{MacDonald:2012mp}. The Post-Newtonian waveform is matched to the
numerical results in the frequency range $f=300-700\,{\rm Hz}$.
Using the tidal phase shift computed in~\cite{damour:12} for the phase difference between the NSBH and BBH Post-Newtonian
waveforms, we find that for our
$C_{\rm NS}=0.156$ neutron star tidal effects would lead to $||\delta h||>1$ for optimally oriented binaries
within a distance $D_{\rm max}\sim 100{\rm Mpc}$ for the Advanced LIGO detector
at design sensitivity. As the signal-to-noise ratio of such an event would be $\sqrt{\langle h,h \rangle}=63.9$, and
given that a signal-to-noise ratio of 8 is required for event detection in Advanced LIGO, this represents about $0.2\%$ 
of the detectable events with these binary parameters (changing the numerical resolution of the NSBH results,
the frequency interval for the matching, and the choice of the Post-Newtonian order used for the tidal dephasing leads to a 
relative uncertainty of $\sim 30\%$ on $D_{\rm max}$, and thus of about a factor of two in the number of events
for which the presence of a neutron star would be detectable). 
Taking into account degeneracies with the mass and spin of
the objects generally reduces this distance by about a factor of $\sim 2-3$~\cite{Lackey:2013}, and thus the event rate by 
at least an order of magnitude. 

These results might seem pessimistic when compared with the predictions obtained for non-spinning systems at an only
slightly lower mass ratio ($q=5$) by Lackey et al~\cite{Lackey:2013}, where the relative error on the tidal
parameter $\Lambda^{1/5}$ 
is measured through a Fisher matrix analysis  ($\Lambda=2/3 k_2 C_{\rm NS}^{-5}$ is the tidal 
deformability and $\Lambda^{1/5}$ is thus mostly proportional to the radius of the neutron star). 
For the neutron star considered here ($\Lambda^{1/5}\sim 3.5$), Lackey et al.~\cite{Lackey:2013} estimate
the relative error in the measurement of $\Lambda^{1/5}$ to be $\left(\sigma_{\Lambda^{1/5}}/\Lambda^{1/5}\right)\sim 0.45$ for an
optimally oriented binary of mass ratio $q=5$ at $100\,{\rm Mpc}$, after marginalizing over the masses of the objects and the spin
of the black hole. This appears to be about 5 times more optimistic than the results presented here.
The two results can, however, easily be reconciled if we remember that the Fisher matrix approximation is based
on a first order expansion of the waveform in the parameters $\bm{\theta}=\{\theta_i\}$ of the binary, i.e.
\beq
\tilde h(\bm{\theta})=\tilde h(\bm{\theta^0}) + \frac{\partial \tilde h}{\partial \theta_i} (\theta_i - \theta_i^0).
\eeq 
The Fisher matrix is then
\beq
\Gamma_{ij}(\bm{\theta^0}) = \langle \frac{\partial \tilde h}{\partial \theta_i}(\bm{\theta^0}),
 \frac{\partial \tilde h}{\partial \theta_j} (\bm{\theta^0})\rangle
\eeq
and the error in the measurement of $\theta_i$ is $\sigma_{\theta_i}=\sqrt{(\Gamma^{-1})_{ii}}$ (see~\cite{Lackey:2013}
for more details). For relative errors of order unity, the choice of expansion parameter can critically
affect the results of the Fisher matrix analysis. In particular, it is easy to see that
\footnote{In Eq.~\ref{eq:siglam}, it is implicitly assumed that $\sigma_{\Lambda^{1/5}}$ is the result
of a Fisher matrix analysis in which $\theta_i=(\Lambda^{1/5},\alpha_k)$ and $\sigma_\Lambda$
the result of a Fisher matrix analysis for $\theta_i=(\Lambda,\alpha_k)$, where the $\alpha_k$
are the same additional parameters in both cases (e.g. masses, black hole spin, coalescence time and
phase,...)}
\beq
\frac{\sigma_{\Lambda^{1/5}}}{\Lambda^{1/5}} = \frac{1}{5} \frac{\sigma_{\Lambda}}{\Lambda},
\label{eq:siglam}
\eeq
a result which is obviously exact for small deviations $\delta \Lambda$ from the true value $\Lambda_0$, as
\beq
(\Lambda_0+\delta\Lambda)^{1/5}=\Lambda_0^{1/5}(1+\delta\Lambda/(5\Lambda_0)+O((\delta\Lambda/\Lambda_0)^2)
\eeq
but not longer holds when $\left|\sigma_{\Lambda^{1/5}}/\Lambda^{1/5}\right|\lo 0.15$. For the configuration
considered in this paper, if the Fisher matrix analysis indicates a relative error of about $45\%$ in a
measurement of $\Lambda^{1/5}$ for an optimally oriented binary at $100\,{\rm Mpc}$, the same analysis
would predict a relative error of $220\%$ in a measurement of $\Lambda$ - a sign that the local
approximation used by the Fisher matrix is no longer valid. Our results indicate that the correct answer is
fairly close to what one would get using a local expansion in $\Lambda$ - which is not too surprising if we
remember that the tidal dephasing $\delta \Psi(f)$ scales linearly with $\Lambda$, and thus $\partial \tilde h
/\partial \Lambda$ is nearly constant as $\Lambda$ varies (while $\partial \tilde h /\partial \Lambda^{1/5}$ vanishes as
$\Lambda \rightarrow 0$). The Fisher matrix analysis expanding in $\Lambda^{1/5}$ does, however, remain perfectly valid
for larger signal-to-noise ratios (i.e. either more advanced detectors, such as the Einstein Telescope, 
closer binaries, or low mass systems). 

\section{Conclusions}

In this work, we compared numerical simulations of binary black hole and neutron star-black hole binaries with
the SpEC code, focusing on a non-spinning binary within the range of mass ratio currently favored by black hole
mass measurements ($M_{\rm BH}=6M_{\rm NS}$). Our results place strong upper limits on the
difference between the two types of binaries in terms of the orbital and merger dynamics, the characteristics of the
remnant black hole, and the gravitational wave signal emitted during the merger for NSBH mergers in which the
neutron star reaches the ISCO without being tidally disrupted. Because of the expected similarity of
the inspiral waveform, these simulations also provide the strongest test so far of the accuracy of the SpEC
code for general relativistic hydrodynamics simulations.

In particular, we measured the rate of periastron advance during the inspiral of the NSBH binary, and found it
to be in good agreement with the results of Le Tiec et al.~\cite{LeTiec-Mroue:2011} for BBH systems. We also showed
that tidal effects in the neutron star are resolved locally in our simulations, and evolve as expected as the 
binary separation shrinks. But, as predicted by Post-Newtonian estimates, they are not significant enough 
to cause measurable differences
in the orbital evolution of the system. In addition to these equilibrium tides, we showed that imperfect initial data
causes the excitation of quadrupolar modes in the neutron star, which ring with little dissipation throughout
the simulation. The amplitude of these oscillations is similar to the initial amplitude of the 
equilibrium tide, but as it does not couple to the orbital quadrupole, nor grow as the binary inspirals, 
its effect on the orbital evolution and waveform should be much smaller than that of the equilibrium tide.

The observable features of the merger and remnant for both types of binaries are found to be nearly impossible to distinguish.
Even during merger, tidal distortion of the neutron star remains minimal, and the final masses and spins of the black
holes are found to be in remarkable agreement, 
to $\delta M_{\rm BH}<5\times 10^{-4}M$ and $\delta \chi_{\rm BH}<0.001$ respectively. 
The velocity kicks, although measured to only $\sim 10\%$ accuracy, are also consistent between the two systems.

Comparisons of the gravitational wave signals show that during inspiral our simulations agree to within a phase 
difference $\delta \phi \sim 0.1-0.2\,{\rm rad}$ (allowing for arbitrary time and phase shifts in the waveforms). 
Considering that these differences dwarf the expected tidal time shift, they represent an independent estimate
of the numerical accuracy of the waveforms. At the time of merger, larger phase errors  $\delta \phi \sim 1\,{\rm rad}$ 
are observed. From the point of view of the Advanced LIGO detector, we find that
the merger waveforms of our BBH and non-disrupting NSBH systems are extremely similar, and would be practically
impossible to differentiate. Surprisingly, it thus appears that even though tidal effects during the inspiral
are small for such a high mass ratio binary, they remain the largest finite size effects measurable by 
ground-based gravitational wave detectors. From Post-Newtonian estimates and considering the Advanced LIGO detector
at design sensitivity we find that, for the $q=6$ non-spinning binary considered here, a $1.4M_\odot$ neutron star
of radius $R_{\rm NS}=13.3\,{\rm km}$ would be impossible to distinguish from a black hole of the same mass for 
any optimally oriented event farther 
than $100\,{\rm Mpc}$, neglecting degeneracies with other binary parameters. 
Once those degeneracies are taken into account, a more realistic requirement would be that the same optimally oriented
event occurs within $\sim 30\,{\rm Mpc}$ of the detector. These predictions are
significantly more pessimistic than results obtained within the Fisher matrix
formalism~\cite{Lackey:2013}, which for some expansion parameters become unreliable when the estimated error on
the size of the neutron star is $\go 15\%$ (although a lot of useful information can
still be obtained from the Fisher matrix results, if their domain of validity is
verified by some direct computations of the differences between waveforms).

From these results, it is quite clear that for non-spinning binaries in the most likely range of black hole masses,
NSBH and BBH binaries cannot be distinguished within the current accuracy of numerical simulations, and are extremely
unlikely to be differentiable in any observable way in the immediate future. As these mergers lack the potential
for post-merger electromagnetic signals (no accretion disk is formed, and no material is ejected from the system), 
and their gravitational wave signals will not be differentiable without significant improvements to
ground-based gravitational wave telescopes, the only hope to prove the presence of a neutron star by another argument
than its measured mass would be pre-merger electromagnetic signals (see e.g.~\cite{TsangEtAl:2012,PalenzuelaEtAl:2013,PaschalidisEtAl:2013}). 
Another consequence of our simulations is that, for the practical purpose of detection
and parameter estimates in Advanced LIGO, NSBH in this part of the parameter space can effectively be modeled
by BBH mergers. While numerical studies of tidal effects and neutron star disruption in lower mass ratio
and/or higher spin systems remain an important task to understand the gravitational wave signals and potential
electromagnetic counterparts of NSBH mergers, there appears to be little immediate incentive for further studies of
low spin NSBH binaries at mass ratio $q\go 5$ with general relativistic hydrodynamics codes, when more accurate solutions
can be obtained at a lower computational cost by solving the binary black hole problem. 

\acknowledgments

The authors wish to thank Tanja Hinderer, Alessandra Buonanno and Andrea Taracchini
for useful discussions regarding tidal effects in NSBH binaries,
the members of the SXS collaboration for their suggestions and support
over the course of this project, Ben Lackey for discussions regarding the detectability
of finite-size effects in gravitational wave signals, and Andy Bohn for comments
on an earlier version of this manuscript.
M.D. acknowledges support through
NASA Grant No.\ NNX11AC37G and NSF Grant PHY-1068243.  H.P., F.F, A.M, I.M.
and M.G. gratefully
acknowledges support from the NSERC of Canada, from the Canada
Research Chairs Program, and from the Canadian Institute for Advanced
Research.  L.K. gratefully acknowledges support from the
Sherman Fairchild Foundation, and from NSF grants PHY-0969111 and
PHY-1005426.  M.S., and B.S. are partially supported by NASA
ATP grant no.\ NNX11AC37G and NSF grants PHY-1151197, PHY-1068881,
and PHY-1005655, by the Sherman Fairchild
Foundation, and the Alfred P. Sloan Foundation. Computations were
performed on the GPC supercomputer at the SciNet HPC
Consortium~\cite{scinet} funded by the Canada Foundation for
Innovation, the Government of Ontario, Ontario Research Fund--Research
Excellence, and the University of Toronto; on Briar{\'e}e from
University of Montreal, under the administration of Calcul Québec and
Compute Canada, supported by Canadian Fundation for Innovation (CFI),
Natural Sciences and Engineering Research Council of Canada (NSERC),
NanoQu{\'e}bec, RMGA and the Fonds de recherche du Qu{\'e}bec - Nature
et technologies (FRQ-NT); and on the Zwicky cluster at Caltech,
supported by the Sherman Fairchild Foundation and by NSF award
PHY-0960291.  

\bibliography{References/References}

\end{document}